\documentclass[12pt]{article} 
\usepackage{mathrsfs,axodraw,amssymb}

\evensidemargin=\oddsidemargin
\def\lag{\mathscr L}
\def\lchi{{\sf P_L}}
\def\rchi{{\sf P_R}}
\def\slash#1{\rlap/#1}
\def\Eq#1{Eq.~(\ref{#1})}
\def\Eqs#1#2{Eqs.~(\ref{#1}) and (\ref{#2})}
\def\fig#1{Fig.~\ref{#1}}
\def\bar{\overline}
\def\tilde{\widetilde}
\def\hat{\widehat}
\def\tilhat#1{\tilde{\anti{#1}}}
\def\pst{{\prime *}}
\def\anti#1{{#1}^c}

\title{\centerline{\normalsize SINP-TNP/06-21 \hfill hep-ph/0608131}
  \bigskip 
\bf Radiative neutrino decay and
     CP-violation in $R$-parity violating supersymmetry}

\author{
Gautam Bhattacharyya$^1$, Palash B. Pal$^1$, Heinrich P\"as$^2$, 
Thomas J. Weiler$^3$ \\ 
\medskip \\ 
\normalsize
$^1${\em Saha Institute of Nuclear Physics, 1/AF Bidhan Nagar, Kolkata
  700064, India}\\   
\normalsize
$^2${\em Department of Physics \& Astronomy, 
University of Hawaii at Manoa,}\\
\normalsize
{\em 2505 Correa Road, Honolulu, HI 96822, USA}\\
\normalsize
$^3${\em Department of Physics and Astronomy, 
Vanderbilt University,}\\
\normalsize {\em Nashville, TN 37235, USA}
}

\date{}

\begin{document}

\maketitle

\begin{abstract}

We calculate the radiative decay amplitude for Majorana neutrinos in
trilinear $R$-parity violating supersymmetric framework.  Our results
make no assumption regarding the masses and mixings of fermions and
sfermions.  The results obtained are exemplary for generic models with
loop-generated neutrino masses.  Comparison of this amplitude with the
neutrino mass matrix shows that the two provide {\em independent\/}
probes of CP-violating phases.
\end{abstract}

\section{Introduction}
In the coventional minimal supersymmetric standard model (MSSM)
without right-handed neutrinos and with conserved $R$-parity it is not
possible to write renormalizable operators which generate neutrino
masses.  However, conservation of lepton and baryon numbers leading to
exact $R$-parity is an {\em ad-hoc} symmetry in the MSSM, and there is
no deep underlying principle as to why such a discrete symmetry will
remain conserved. Once we admit $R$-parity violating (RPV) couplings
\cite{rpar} in the superpotential, neutrino masses are generated
through diagrams involving RPV couplings.  These are Majorana
masses which result as a consequence of broken lepton number(s). There
are two kinds of explicit RPV couplings: bilinear and trilinear. The
bilinear ones induce sneutrino vacuum expectation values, allowing
neutrinos to mix with the neutralinos, and in this mechanism only one
physical neutrino obtains mass at the tree level \cite{bilinear}.
Trilinear Yukawa type couplings generate a complete neutrino mass
matrix through one-loop self-energy graphs \cite{trilinear}.
Obviously, one can take a combination of bilinear and trilinear
couplings, along with possible soft terms, for a complete analysis
\cite{abl}.  Here we concentrate only on the trilinear superpotential
couplings.  Importantly, the diagrams that generate neutrino masses
also generate dipole moments by insertion of photons to the internal
lines. Clearly, in the absence of Dirac-type operators, such dipole 
moments are transitional.

The connection between neutrino Majorana masses and transition moments
have been studied in the past \cite{babu_moha,barbieri} and
predictions have been made using the constraints on trilinear RPV
couplings placed from the data on neutrino masses and mixings
\cite{bkp} (compare the present update in \cite{faes}).  However, as
regards the analytic formulation of dipole moments, two crucial things
were still missing.  One of these is a general analysis with the
inclusion of fermion and sfermion flavor mixings.  The other is a
careful handling of the phases from different sources that enter the
analysis.  We try to address these two issues in the present paper.
In doing so, we ask the following question: how different is the
combination of phases that appears in the neutrino radiative decay
from the one that appears in the neutrinoless double beta decay?  We
keep the formalism at a very general level.  Thus our results can be
applied, with appropriate changes in notation, to a general class of
models \cite{general} involving Majorana neutrinos, although for the
sake of definiteness we choose the supersymmetric RPV model, for which
the lepton number violating part of the RPV superpotential can be
written as
\begin{eqnarray}
\mathscr W = {1\over 2} \sum_{ijk} \lambda_{ijk} L_i L_j \anti E_k + 
\sum_{ijk} \lambda'_{ijk} L_i Q_j \anti D_k \,.
\label{supot}
\end{eqnarray}
Here $L_i$ denotes the doublet leptonic superfield and $Q_i$ is the
doublet quark superfield, whereas $\anti E_i$ and $\anti D_i$ denote
SU(2)-singlet superfields which contain the left-chiral components of
charged antileptons and down-type antiquarks.  Stringent upper limits
exist on all these couplings from different experiments
\cite{review,gb95}.

\section{Feynman rules with fermion and sfermion mixings}
Without any loss of generality, we can take the quark superfields in a
basis such that the up-type quark mass matrix will come out to be
diagonal.  Also, the lepton superfields will be defined in a basis
where the charged lepton mass matrices will come out to be diagonal.
In addition, we can take the singlet fields in a way that they will
not have to be diagonalized further.  This will be called the {\em
flavor basis}.

The mass matrices for neutrinos and the $d$-type quarks will be
non-diagonal in this basis.  We define the mass eigenstates as
follows:
\begin{eqnarray}
\nu_{iL} &=& \sum_\alpha U_{i\alpha} \nu_{\alpha L} \,,
\nonumber\\* 
d_{i L} &=& \sum_a V_{ia} d_{aL} \,,
\label{fmix}
\end{eqnarray}
where in each case, the states on the left side are the flavor states,
and the states on the right are mass eigenstates.  The matrices $U$
and $V$ are respectively the PMNS and the CKM mixing matrices for
leptons and quarks.  The squark fields will also mix in general.  We
will denote the superpartners of the left-chiral quark fields $d_L$ by
$\tilde d$, without the subscript $L$ since the scalar field does not
carry any chirality.  Similarly $\anti d_L$, the conjugate of
right-chiral quark fields $d_R$, have superpartners which will be
denoted by $\tilde{\anti d}$.  In the mass matrix, the fields $\tilde
d$ will mix with the fields $\tilde{\anti d}^\dagger$.  For three
generations, there will be then six eigenstates for down squarks.
These will be denoted by $\Delta_A$, where
\begin{eqnarray}
\tilde d_k &=& \sum_A K_{kA} \Delta_A \,, \nonumber\\*
\tilde {\anti d_k}^\dagger &=& \sum_A \hat K_{kA} \Delta_A \,.
\label{defK}
\end{eqnarray}
Here, for $n$ fermion generations, both $K$ and $\hat K$ are $n\times
2n$ matrices, which are respectively the first and last $n$ rows of a
$2n\times 2n$ unitary mixing matrix $\mathbb K$ of the down squark
sector.

From the superpotential given in \Eq{supot}, when we derive the
Lagrangian containing ordinary fields, any cubic term in the
superpotential produces various kinds of Yukawa couplings.  We are
interested in processes with external neutrino lines, so we seek for
terms where the neutrino field (and not the sneutrino field) appears.
For the time being, let us deal with only the second term, i.e., the
$\lambda'$ term.  We shall comment on the inclusion of the $\lambda$
term later.  The relevant Yukawa couplings are then:
\begin{eqnarray}
\lag_Y' &=& 
\sum_{ijk} \lambda'_{ijk} \nu_{iL}^\top C^{-1} d_{jL} \tilhat
d_k + 
\sum_{ijk} \lambda'_{ijk} \nu_{iL}^\top C^{-1} \anti d_{kL} \tilde
d_j + \mbox{h.c.} \,.
\end{eqnarray}
Using \Eq{fmix}, this can be rewritten in terms of mass eigenstates
in the form
\begin{eqnarray}
\lag_Y' &=& 
\sum_{\alpha aA} \hat \Lambda'_{\alpha aA} 
\nu_\alpha^\top C^{-1} \lchi d_a \Delta_A^\dagger + 
\sum_{\alpha a A} \Lambda'_{\alpha aA} 
\nu_\alpha^\top C^{-1} \lchi \anti d_a \Delta_A + \mbox{h.c.}\,,
\label{LY'}
\end{eqnarray}
where 
\begin{eqnarray}
\hat \Lambda'_{\alpha aA} &=& \sum_{ijk} \lambda'_{ijk} U_{i\alpha}
V_{ja} \hat K_{kA}^* \,, \nonumber\\*
\Lambda'_{\alpha aA} &=& \sum_{ijk} \lambda'_{ijk} U_{i\alpha}
\delta_{ka} K_{jA} \,.
\label{Lambda'}
\end{eqnarray}
The appearance of the Kronecker delta in the latter formula is a
reminder that for singlet fields, the flavor indices and the mass
indices are interchangeable.  In \Eq{LY'}, we have also used the
notation for chiral projection operators that will be used henceforth:
\begin{eqnarray}
\lchi = \frac12 (1-\gamma_5) \,, \qquad
\rchi = \frac12 (1+\gamma_5) \,.
\end{eqnarray}

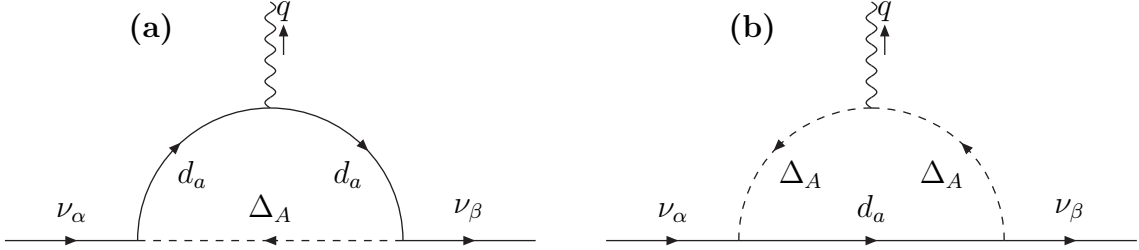
\begin{figure}[t]
\begin{center}
\begin{picture}(200,100)(-100,0)
\ArrowLine(-100,0)(-50,0)
\Text(-75,7)[b]{$\nu_\alpha$}
\ArrowLine(50,0)(100,0)
\Text(75,7)[b]{$\nu_\beta$}
\DashArrowLine(50,0)(-50,0)3
\Text(0,7)[b]{$\Delta_A$}
\ArrowArcn(0,0)(50,180,90)
\Text(-35,25)[l]{$d_a$}
\ArrowArcn(0,0)(50,90,0)
\Text(35,25)[r]{$d_a$}
\Photon(0,50)(0,90)25
\LongArrow(5,70)(5,80)
\Text(5,83)[b]{$q$}
\Text(-45,80)[]{\bf (a)}
\end{picture}
\qquad
\begin{picture}(200,100)(-100,0)
\ArrowLine(-100,0)(-50,0)
\Text(-75,7)[b]{$\nu_\alpha$}
\ArrowLine(50,0)(100,0)
\Text(75,7)[b]{$\nu_\beta$}
\ArrowLine(-50,0)(50,0)
\Text(0,7)[b]{$d_a$}
\DashArrowArc(0,0)(50,90,180)3
\Text(-35,25)[l]{$\Delta_A$}
\DashArrowArc(0,0)(50,0,90)3
\Text(35,25)[r]{$\Delta_A$}
\Photon(0,50)(0,90)25
\LongArrow(5,70)(5,80)
\Text(5,83)[b]{$q$}
\Text(-45,80)[]{\bf (b)}
\end{picture}
\end{center}
\caption{1-loop diagrams for the neutrino photon vertex.  All fields
  shown are mass eigenstates.  The photon attaches to the quark in
  diagram (a), and to the squark in (b).}\label{f:dir} 
\end{figure}

Contributions to radiative neutrino decay occur at the 1-loop level
through diagrams such as those shown in \fig{f:dir}.  If the matrix
$\mathbb K$ is block-diagonal, i.e., there is no mixing between the
fields $\tilde d$ and the fields $\tilde{\anti d}^\dagger$ in the mass
matrix, contributions will be proportional to the external neutrino
masses and should therefore be small.  We will be interested in the
more general case where $\mathbb K$ is not necessarily block-diagonal.
Diagrams for this general case are shown in \fig{f:dir}.  As we will
see, in this case the radiative neutrino decay amplitude will be
proportional to the masses of the down-type quarks.

Before calculating the amplitude, let us put \Eq{LY'} is a more
conventional form.  We have used the symbol $\anti d$.  This is the
conjugate of the $d$-quark field.  For any fermion field $\psi$, the
Lorentz-covariant conjugate can be defined as
\begin{eqnarray}
\anti\psi = \gamma_0 C \psi^* \,.
\label{defanti}
\end{eqnarray}
It is then easy to see that
\begin{eqnarray}
\psi^\top C^{-1} = \psi^\top C^\dagger 
= \Big( C\psi^* \Big)^\dagger =  \Big( \gamma_0 \anti\psi
\Big)^\dagger = \overline{\anti\psi} \,.
\label{psitcinv}
\end{eqnarray}
Antisymmetry of the matrix $C$ follows from the relation
$\anti{(\anti\psi)} = \psi$.   Using this as well as the
anticommutation of fermion fields, we can write
\begin{eqnarray}
\nu_\alpha^\top C^{-1} \lchi \anti d_a = {\anti d_a}^\top C^{-1} \lchi
\nu_\alpha = \bar d_a \lchi \nu_\alpha \,,
\label{nuLdc}
\end{eqnarray}
using \Eq{psitcinv} in the last step.

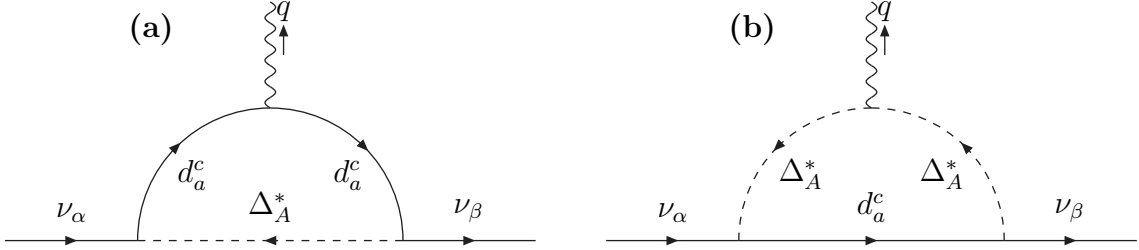
\begin{figure}[t]
\begin{center}
\begin{picture}(200,100)(-100,0)
\ArrowLine(-100,0)(-50,0)
\Text(-75,7)[b]{$\nu_\alpha$}
\ArrowLine(50,0)(100,0)
\Text(75,7)[b]{$\nu_\beta$}
\DashArrowLine(50,0)(-50,0)3
\Text(0,7)[b]{$\Delta_A^*$}
\ArrowArcn(0,0)(50,180,90)
\Text(-35,25)[l]{$\anti d_a$}
\ArrowArcn(0,0)(50,90,0)
\Text(35,25)[r]{$\anti d_a$}
\Photon(0,50)(0,90)25
\LongArrow(5,70)(5,80)
\Text(5,83)[b]{$q$}
\Text(-45,80)[]{\bf (a)}
\end{picture}
\qquad
\begin{picture}(200,100)(-100,0)
\ArrowLine(-100,0)(-50,0)
\Text(-75,7)[b]{$\nu_\alpha$}
\ArrowLine(50,0)(100,0)
\Text(75,7)[b]{$\nu_\beta$}
\ArrowLine(-50,0)(50,0)
\Text(0,7)[b]{$\anti d_a$}
\DashArrowArc(0,0)(50,90,180)3
\Text(-35,25)[l]{$\Delta_A^*$}
\DashArrowArc(0,0)(50,0,90)3
\Text(35,25)[r]{$\Delta_A^*$}
\Photon(0,50)(0,90)25
\LongArrow(5,70)(5,80)
\Text(5,83)[b]{$q$}
\Text(-45,80)[]{\bf (b)}
\end{picture}
\end{center}
\caption{Same as in \fig{f:dir}, except the internal lines are now
  conjugated.}\label{f:conj}
\end{figure}

Further, for Majorana neutrino fields, the conjugate field is the same
as the original fields, apart from possibly a phase factor:
\begin{eqnarray}
\nu_\alpha = \eta_\alpha \anti\nu_\alpha \,.
\label{majo}
\end{eqnarray}
Thus, for Majorana neutrino fields, we obtain
\begin{eqnarray}
\nu_\alpha^\top C^{-1} = \overline{\anti\nu}_\alpha =
\eta_\alpha \overline\nu_\alpha \,.
\label{nubar}
\end{eqnarray}
Putting \Eqs{nuLdc}{nubar} back into \Eq{LY'}, we obtain
\begin{eqnarray}
\lag_Y' = 
\sum_{\alpha aA} \bigg[ \bar\nu_\alpha \Big( \eta_\alpha \hat
\Lambda'_{\alpha aA} \lchi + \Lambda^\pst_{\alpha aA} \rchi \Big) d_a
\Delta_A^\dagger + 
\bar d_a \Big( \eta_\alpha^* \hat
\Lambda^\pst_{\alpha aA} \rchi + \Lambda'_{\alpha aA} \lchi \Big)
\nu_\alpha \Delta_A \bigg] \,,
\label{L'Yall}
\end{eqnarray}
where we have written the hermitian conjugate terms explicitly. By
using \Eqs{defanti}{majo}, these interactions can also be written as
\begin{eqnarray}
\lag_Y' = 
\sum_{\alpha aA} \bigg[ \bar\anti d_a \Big( \hat
\Lambda'_{\alpha aA} \lchi + \eta_\alpha ^* \Lambda^\pst_{\alpha aA}
\rchi \Big) \nu_\alpha \Delta_A^\dagger + 
\bar \nu_\alpha \Big( \hat
\Lambda^\pst_{\alpha aA} \rchi + \eta_\alpha \Lambda'_{\alpha aA}
\lchi \Big) 
\anti d_a \Delta_A \bigg] \,.
\label{L'Yconj}
\end{eqnarray}
This form will be useful when we calculate the diagrams shown in
\fig{f:conj}, where the internal particles are conjugated.  As is
well-known, such cojugated diagrams exist when we deal with Majorana
neutrinos~\cite{Pal:1981rm}.

\section{Radiative neutrino decay amplitude}
\subsection{Calculation of the vertex function}
The vertex function $\Gamma^\lambda$ will be defined in such a way
that the matrix element of the photon vertex with on-shell fermions
is given by
\begin{eqnarray}
{\cal M} = \bar u_\beta (p') \Gamma^\lambda u_\alpha (p)
A_\lambda (q) \,.
\label{emvertex}
\end{eqnarray}

For the diagram in \fig{f:dir}a, the contribution to the vertex
function will be denoted by $\Gamma^\lambda_{(\ref{f:dir}a)}$, which
can be easily written down, using the vertex Feynman rules from the
interaction given in \Eq{L'Yall}.  The result is
\begin{eqnarray}
i\Gamma^\lambda_{(\ref{f:dir}a)} &=& \sum_{aA} \int {d^4k \over (2\pi)^4} 
i\Big( \eta_\beta
\hat \Lambda'_{\beta aA} \lchi + \Lambda^\pst_{\beta aA} \rchi \Big) 
{i(\slash k - \slash q + m_{d_a}) \over
(k-q)^2 - m_{d_a}^2} \nonumber\\*  
&& \qquad \times \Big( ie_d \gamma^\lambda \Big) 
{i(\slash k+ m_{d_a}) \over k^2 - m_{d_a}^2} 
i\Big( \eta_\alpha^* \hat
\Lambda^\pst_{\alpha aA} \rchi + \Lambda'_{\alpha aA} \lchi \Big)
{i \over (k-p)^2 - \tilde M_{\Delta_A}^2} \,,
\label{gam1}
\end{eqnarray}
where $e_d$ is the charge of the $d$-quark.  From a general analysis
of the form factors related to the electromagnetic vertices, we know
that the radiative transition amplitudes can have only the magnetic
and electric dipole form factors.  These are helicity flipping terms,
containing an even number of Dirac matrices sandwiched between the
spinors.  We can separate out such terms and write
\begin{eqnarray}
\Gamma^\lambda_{(\ref{f:dir}a)} &=& ie_d \sum_{aA} m_{d_a} 
\Big( \eta_\beta
\hat \Lambda'_{\beta aA} \Lambda'_{\alpha aA} \lchi 
+ \eta_\alpha^* \hat
\Lambda^\pst_{\alpha aA} \Lambda^\pst_{\beta aA} \rchi \Big)
\nonumber\\* 
&& \times 
\int {d^4k \over (2\pi)^4} 
{(\slash k - \slash q) \gamma^\lambda + \gamma^\lambda \slash k \over
  \Big[k^2 - m_{d_a}^2\Big] \Big[(k-q)^2 - m_{d_a}^2\Big] \Big[(k-p)^2
    - \tilde M_{\Delta_A}^2\Big]} \,. 
\end{eqnarray}
Similarly, we can write down the contribution of \fig{f:dir}b.  
By combining the factors in the denominator of the integrand, we can
identify the magnetic and electric dipole terms, which are of the form 
\begin{eqnarray}
\Gamma^\lambda_{(\ref{f:dir})} &=& ie_d 
\sigma^{\lambda\rho}q_\rho  
\sum_{aA} m_{d_a} 
\Big( \eta_\beta
\hat \Lambda'_{\beta aA} \Lambda'_{\alpha aA} \lchi 
+ \eta_\alpha^* \hat
\Lambda^\pst_{\alpha aA} \Lambda^\pst_{\beta aA} \rchi \Big)
J_{aA} \,.
\label{defJ}
\end{eqnarray}
Contributions to $J_{aA}$ from \fig{f:dir}a and \fig{f:dir}b are given
by
\begin{eqnarray}
J_{aA}^{(\ref{f:dir}a)} = i \int_0^1 dx \int {d^4k \over (2\pi)^4}
{-2 (1-x)^2 \over \Big[ k^2 - x \tilde M_{\Delta_A}^2 - (1-x) m_{d_a}^2
    \Big]^3} \,, 
\label{Jsimp} \\
J_{aA}^{(\ref{f:dir}b)} = i \int_0^1 dx \int {d^4k \over (2\pi)^4}
{-2 x(1-x) \over \Big[ k^2 - x \tilde M_{\Delta_A}^2 - (1-x) m_{d_a}^2
    \Big]^3} \,. 
\label{Jb}
\end{eqnarray}
In writing these integrals, we have neglected neutrino masses in the
denominators.  Thus
\begin{eqnarray}
J_{aA}^{(\ref{f:dir})} = i \int_0^1 dx \int {d^4k \over (2\pi)^4}
{-2 (1-x) \over \Big[ k^2 - x \tilde M_{\Delta_A}^2 - (1-x) m_{d_a}^2
    \Big]^3} \,.
\end{eqnarray}

We now come to the conjugated diagrams of \fig{f:conj}, which are
related to the diagrams of \fig{f:dir} through 
\begin{eqnarray}
d_a \longrightarrow \anti d_a \,, \qquad \Delta_A \longrightarrow
\Delta_A^\dagger \,.
\end{eqnarray}
It is now convenient to use the Yukawa couplings in the form given in
\Eq{L'Yconj}.  Comparing this with \Eq{L'Yall}, we see that the
corresponding amplitudes are related by the substitutions
\begin{eqnarray}
\Lambda_{\alpha aA} \longleftrightarrow \hat
\Lambda_{\alpha aA} \,,
\end{eqnarray}
and an overall negative sign because the photon line now attaches to
an internal line of opposite charge.  So we obtain
\begin{eqnarray}
\Gamma^\lambda_{(2)} &=& \null - ie_d \sigma^{\lambda\rho}q_\rho 
\sum_{aA} m_{d_a} 
\Big( \eta_\beta
\hat \Lambda'_{\alpha aA} \Lambda'_{\beta aA} \lchi 
+ \eta_\alpha^* \hat
\Lambda^\pst_{\beta aA} \Lambda^\pst_{\alpha aA} \rchi \Big)
J_{aA} \,.
\end{eqnarray}
Combining the contributions of
Figs.~\ref{f:dir} and \ref{f:conj}, we can therefore write the total 
vertex function $\Gamma^\lambda$ as
\begin{eqnarray}
\Gamma^\lambda = ie_d \sigma^{\lambda\rho}q_\rho 
\sum_{aA} m_{d_a} \bigg[ 
\eta_\beta \Big(
\hat \Lambda'_{\beta aA} \Lambda'_{\alpha aA} 
- \hat \Lambda'_{\alpha aA} \Lambda'_{\beta aA} 
\Big) \lchi \quad \quad \nonumber\\ 
+ \eta_\alpha^* \Big( 
\hat\Lambda^\pst_{\alpha aA} \Lambda^\pst_{\beta aA} 
- \hat\Lambda^\pst_{\beta aA} \Lambda^\pst_{\alpha aA} 
\Big) \rchi \bigg] J_{aA} \,.
\label{Gamma}
\end{eqnarray}
Clearly, if $\alpha=\beta$, i.e., we are talking about the diagonal
vertex, the contribution vanishes, as is required by CPT invariance.

\subsection{CP-invariant limit}
Before proceeding further with the calculation, it would be
instructive to check the form of the vertex in the CP-invariant
limit.  The CP transformation property of a fermion field $\psi$ is
given by
\begin{eqnarray}
(\mathscr{CP}) \psi(\vec x, t) (\mathscr{CP})^{-1} = \xi^* C
\psi^* (-\vec x, t) \,,
\end{eqnarray}
where $\xi$ can be called the CP phase of the field.  For the scalar
and pseudoscalar bilinears that occur in the Yukawa coupling of
\Eq{L'Yall}, the effect of CP transformation is then given by
\begin{eqnarray}
(\mathscr{CP}) \bar\psi_1 \psi_2 (\mathscr{CP})^{-1} &=& 
\xi_1 \xi_2^* \; \bar \psi_2 \psi_1 \,, \nonumber
\\*
(\mathscr{CP}) \bar\psi_1 \gamma_5 \psi_2 (\mathscr{CP})^{-1} 
&=& - \xi_1 \xi_2^* \; \bar \psi_2 \gamma_5 \psi_1 \,,
\end{eqnarray}
where the space-time points have not been explicitly mentioned.  For
all Dirac fermion fields, the CP phase can be absorbed into the
definition of the antiparticle without any loss of generality.  For
Majorana neutrinos, however, that cannot be done because we have
already defined $\anti\nu_\alpha$ through \Eq{majo}.  In what follows,
we will denote the CP phase of the field $\nu_\alpha$ by $\xi_\alpha$.

We now take the CP transformation of the first term on the right side
of \Eq{L'Yall}.  
\begin{eqnarray}
(\mathscr{CP}) \bar\nu_\alpha \Big( \eta_\alpha \hat
\Lambda'_{\alpha aA} \lchi + \Lambda^\pst_{\alpha aA} \rchi \Big) d_a
\Delta_A^\dagger (\mathscr{CP})^{-1} = \xi_\alpha \bar d_a \Big(
\eta_\alpha \hat \Lambda'_{\alpha aA} \rchi + \Lambda^\pst_{\alpha aA}
\lchi \Big) \nu_\alpha \Delta_A \,.
\end{eqnarray}
If the Lagrangian is CP invariant, this CP transform should equal the
hermitian conjugate term present in \Eq{L'Yall}.  This imposes the
following conditions on the Yukawa couplings:
\begin{eqnarray}
\hat \Lambda^\pst_{\alpha aA} &=& \xi_\alpha \eta_\alpha^2
\hat \Lambda'_{\alpha aA} \,, \nonumber \\*
\Lambda^\pst_{\alpha aA} &=& \xi_\alpha^* \Lambda'_{\alpha aA} \,.
\label{CPcond}
\end{eqnarray}
If we put these conditions back into \Eq{Gamma}, we find that the
expression for the vertex function reduces to the form
\begin{eqnarray}
\Gamma^\lambda 
= ie_d \sigma^{\lambda\rho}q_\rho \eta_\beta \sum_{aA} m_{d_a} \bigg[ 
\hat \Lambda'_{\beta aA} \Lambda'_{\alpha aA} \Big( \lchi 
- \zeta_{\alpha\beta}^* \rchi \Big) 
+ \hat \Lambda'_{\alpha aA}
\Lambda'_{\beta aA} \Big( 
\zeta_{\alpha\beta} \rchi - \lchi \Big) 
\bigg] 
J_{aA} \,,
\label{GammaCP}
\end{eqnarray}
where
\begin{eqnarray}
\zeta_{\alpha\beta} = \xi_\alpha \eta_\alpha \xi_\beta^* \eta_\beta^*
= {\xi_\alpha \eta_\alpha \over \xi_\beta \eta_\beta} \,.
\end{eqnarray}
For Majorana fields, the quantity $\xi\eta$ equals the CP eigenvalue
of the 1-particle states 
\cite{Kayser:1983wm, Kayser:1984ge, Schechter:1981hw}, so that
$\zeta_{\alpha\beta}$ denotes the relative CP eigenvalues of the two
Majorana neutrinos involved in the process.  Thus, \Eq{GammaCP} shows
that in the CP-conserving case, the transition amplitude is purely
electric dipole type if the CP eigenvalues are the same for the two
neutrinos, and purely magnetic dipole type if the CP eigenvalues are
opposite.  This is expected on quite general grounds
\cite{Nieves:1981zt}.

\subsection{CP-violating scenario}
We now go back to the general case, where the couplings do not
necessarily obey \Eq{CPcond} and therefore the Lagrangian does not
conserve CP.  For such a case, we cannot use the vertex of
\Eq{GammaCP}.  Rather, we need to go back to the general form of the
vertex given in \Eq{Gamma}.  In fact, this expression can be written
in a simpler form, noting the fact that the neutrino mixing matrix $U$
and the phases $\eta_\alpha$ are interconnected.  In general, any mass
matrix $M$ can be diagonalized by a bi-unitary transformation
involving two unitary matrices $U$ and $U'$:
\begin{eqnarray}
U'^\dagger M U = D \,,
\label{biunitary}
\end{eqnarray}
where $D$ is a diagonal matrix whose diagonal elements are the mass
eigenvalues.  \Eq{biunitary} implies $U^\top M^\top U'^* = D^\top$.
For Majorana neutrinos, the mass matrix is symmetric.  So, using
$M=M^\top$ and $D=D^\top$, we see that it is possible to {\em
choose\/} $U'=U^*$.  Thus the bi-unitary transformation reduces to 
\begin{eqnarray}
U^\top M U = D \,,
\label{UMU}
\end{eqnarray}
We will adopt this convention in what follows.  This automatically
implies~\cite{Mohapatra:1998rq}
\begin{eqnarray}
\eta_\alpha = 1 \qquad \mbox{for each $\alpha$.}
\label{eta=1}
\end{eqnarray}
Of course, other choices are possible \cite{Mohapatra:1998rq}, but
they all lead to same physical results.  Therefore, the choice of
\Eq{eta=1} does not mean any loss of generality.

To proceed, we make some simplifying assumptions.  First, motivated by
the fact that quark mixing is small, we will neglect it altogether and
use the unit matrix in place of $V$.  As mentioned earlier, the
$\tilde d$-$\tilde{\anti d}^\dagger$ mixing is crucial for the terms
that we have been considering in Figs. \ref{f:dir} and \ref{f:conj},
viz., terms having a factor of a down-type quark mass, let us assume
that these are the only kind of mixings in the squark sector mass
matrix.  Further we assume that there is no intergenerational mixing
in this sector, which means that the $\tilde d$ squark of a certain
generation mixes only with $\tilde{\anti d}^\dagger$ of the same
generation.  Then the mixing matrix $\mathbb K$ in the squark sector
is real, and is given by
\begin{eqnarray}
\mathbb K \equiv \left( \begin{array}{c}K \\ \hat K\end{array} \right)
= \left( \begin{array}{cc} C & S \\  -S & C\end{array} 
\right) \,,
\end{eqnarray}
where in the final form, each entry denotes a $n\times n$ diagonal
matrix, where $n$ is the number of generations, and
\begin{eqnarray}
C \equiv \mbox{diag}\; (\cos\theta_1, \cos\theta_2, \cdots), \quad
S \equiv \mbox{diag}\; (\sin\theta_1, \sin\theta_2, \cdots) .
\end{eqnarray}
More explicitly, we can write the elements of $K$ and $\hat K$ as
\begin{eqnarray}
K_{kA} &=& \cos\theta_k \delta_{k,A} + \sin\theta_k \delta_{k+n,A}
\nonumber \\* 
\hat K_{kA} &=& -\sin\theta_k \delta_{k,A} + \cos\theta_k
\delta_{k+n,A} \,. 
\end{eqnarray}
It should be recalled, from the definition of \Eq{defK}, that the
index $A$ runs over $2n$ values for $n$ generations.

The only intergenerational mixing in this picture is in the neutrino
sector, which contains the phases of our concern.  The vertex of
\Eq{Gamma} can then be written as:
\begin{eqnarray}
\Gamma^\lambda = ie_d \sigma^{\lambda\rho}q_\rho \Big[ 
F \lchi + F^* \rchi \Big] \,,
\label{GammaF}
\end{eqnarray}
where
\begin{eqnarray}
F = \sum_{jk} m_{d_j} B_{jk} \cos\theta_k \sin\theta_k \Big( 
J_{j,k+n} - J_{jk} \Big)\,,
\label{F}
\end{eqnarray}
with $B_{jk}$ defined by
\begin{eqnarray}
B_{jk} = \sum_{ii'} \Big( U_{i'\alpha} U_{i\beta} - U_{i\alpha}
U_{i'\beta} \Big) \lambda'_{ijk} \lambda'_{i'kj} \,.
\label{B}
\end{eqnarray}

Apparently, there is one big difference between this form of the
vertex function and the expressions derived in an earlier work on the
subject \cite{bkp}.  The expression in Ref.~\cite{bkp} contains two
explicit factors of down quark masses.  It should be emphasized that
the extra factor, compared to our \Eq{F}, appears from an extra
assumption.  To appreciate this, consider what happens if all
down-type squarks were really degenerate.  The integrals defined in
\Eqs{Jsimp}{Jb} would have been independent of the squark index $A$ in
that case.  Looking at \Eq{Lambda'}, we find that in \Eq{Gamma}, the
sum over $A$ would have involved only the combination
\begin{eqnarray}
\sum_A \hat K_{kA}^* K_{jA} = 0 \,,
\label{KK}
\end{eqnarray}
because of the orthogonality of two different rows of the unitary
matrix $\mathbb K$.  The result implies that the vertex has to involve
only differences of squark masses.  These differences have to be small
from various phenomenological requirements, and one often assumes that
the differences are proportional to the corresponding quark masses.
In particular, \Eq{F} shows that in absence of intergenerational
mixing, the contribution from the two squarks of a single generation
vanishes if they are degenerate.  Customarily, one assumes that the
mass squared differences of the two squarks in the same generation is
of the form
\begin{eqnarray}
\Delta \tilde M^2 = \tilde M m_q \,,
\end{eqnarray}
where $m_q$ is the mass of the quark in the same generation and
$\tilde M$ is the average mass of the two squarks.  It is this
assumption which produced the extra factor of quark mass in the
expression of Ref.~\cite{bkp}.  We will carry on our discussion
without this specific form for the squark mass differences.  The
phenomenological consequences of the decay rate, and the resulting
bounds obtained thereby, have been discussed in earlier papers
\cite{bkp,faes}.  In our discussion, we highlight CP violating
features of this amplitude.

\section{Exploring phenomenological consequences}
\subsection{The mass matrix}
Since the mixing matrix $U$ is present in the vertex function through
the combination of \Eq{B}, the radiative decay amplitudes will clearly
involve CP violating phases.  To analyze how they appear in the
amplitude, we first look into the generation of neutrino mass and
mixing in this model.  In the flavor basis for the neutrinos, the mass
terms can be written in the form
\begin{eqnarray}
\sum_{i,i'}\bar\nu_i \Big( M_{ii'} \lchi + M_{i'i}^* \rchi \Big)
\nu_{i'} \,,
\end{eqnarray}
where the matrices appearing in the two chiral terms are related by
the condition of hermiticity of their sum.  Because of this relation,
we can work entirely with just one part, say $M$, and call it the mass
matrix.  Diagonalization of this matrix was discussed in
\Eqs{biunitary}{UMU}.  Once this part is diagonalized and the 
eigenbasis is determined for left- and right-chiral
fields, the other term is automatically diagonal in
the same basis.

It is to be noted that trilinear RPV couplings do not generate
neutrino masses at the tree level.  The masses are generated through
1-loop diagrams which look like the diagrams of Figs.~\ref{f:dir} and
\ref{f:conj} {\em without\/} the photon line.  If we calculate these
mass diagrams with neutrino flavor states on the outer lines and with
vanishing external momentum, we obtain
\begin{eqnarray}
M_{ii'} = \sum_{aA} m_{d_a} 
\left( \hat h'_{iaA} h'_{i'aA}  
+ \hat h'_{i'aA} h'_{iaA}  \right) J'_{aA} \,,
\end{eqnarray}
with the integral $J'_{aA}$ defined as
\begin{eqnarray}
J'_{aA} = i \int {d^4k \over (2\pi)^4} 
{1 \over [k^2 - m_{d_a}^2] [k^2 - \tilde M_{\Delta_A}^2]} \,,
\end{eqnarray}
and the couplings defined by 
\begin{eqnarray}
\hat h'_{iaA} &=& \sum_{jk} \lambda'_{ijk} 
V_{ja} \hat K_{kA}^* \,, \nonumber\\*
h'_{iaA} &=& \sum_{jk} \lambda'_{ijk} \delta_{ka} K_{jA} \,,
\end{eqnarray}
which are reminiscent of the couplings $\hat\Lambda$ and $\Lambda$ in
the neutrino flavor basis.  The integral $J'$ itself is
logarithmically divergent, but we need to remember that the argument
around \Eq{KK} guarantees that $M_{ii'}$ vanishes if all squarks are
degenerate.  Since the divergence is independent of all masses, it
cancels anyway.  Of course, this has to happen since the theory is
renormalizable and there is no tree-level mass term.

In the simplified scenario advocated above where we neglect quark
mixing and intergenerational squark mixing as well, the mass formula
trivially reduces to
\begin{eqnarray}
M_{ii'} &=& \sum_{jk} m_{d_j} 
\left( \lambda'_{ijk} \lambda'_{i'kj}  
+ \lambda'_{i'jk} \lambda'_{ikj}  \right) \sin\theta_k \cos\theta_k
\Big( J'_{j,k+n} - J'_{jk} \Big) \,, 
\label{M}
\end{eqnarray}
which is what we will use below.

\subsection{Ansatz for a simplified analysis}\label{s:ans}
Phases can enter the mass matrix from any of the couplings
$\lambda'_{ijk}$.  For three generations, there are 27 such couplings,
and the phases are all independent.  For illustrative purpose, we
make a simplifying assumption that the phases can be written in the
form
\begin{eqnarray}
\lambda'_{ijk} = e^{i(\omega_i+\phi_j+\chi_k)} \Big| \lambda'_{ijk}
\Big| \,, \qquad \forall i,j,k \,.
\label{ansatz}
\end{eqnarray}

Since we are considering leptonic processes, it would have been most
natural to focus on the phases $\omega_i$ only, which pertain to
lepton doublets.  However, it is easy to see that if these are the
only RPV couplings, i.e., the couplings are of the form
\begin{eqnarray}
\lambda'_{ijk} = e^{i\omega_i} \Big| \lambda'_{ijk} \Big| \,,
\qquad \forall i,j,k \,,
\label{omega}
\end{eqnarray}
the model conserves CP.  The reason is that in this case, we can
absorb all phases of the $\lambda'_{ijk}$'s into the definition of the
leptonic superfields $L_i$.  The phases of these superfields will then
reappear in the Yukawa couplings with the relevant Higgs doublet
superfield, but those couplings can again be made real by readjusting
the phases of the superfields $\anti E_j$.  Thus, at the end, no CP
violating phase remains.  If the phases denoted by $\phi$ and $\chi$
are non-zero, the possibility of CP violation remains, but the phases
denoted by $\omega$ continue to be irrelevant for the reason stated
above.

So we drop $\omega_i$ altogether.  \Eq{ansatz} now boils down to
\begin{eqnarray}
\lambda'_{ijk} = e^{i(\phi_j + \chi_k)} \Big| \lambda'_{ijk} \Big| \,,
\qquad \forall i,j,k \,.
\label{quarkansatz}
\end{eqnarray}
Through \Eq{M}, this will introduce phases in the neutrino mass
matrix.  This mass matrix is diagonalized by the matrix $U$ through
\Eq{UMU}, so that the phases in $U$ are functions of the $\phi$'s and
the $\chi$'s.  The functional relation is, however, non-trivial for
three generations, where there are two Majorana-type and one
Dirac-type phases in the neutrino mixing matrix~\cite{Doi:1980yb}.

We, however, note from the expressions in \Eqs{F}{B} that phases can
enter the amplitude only through the quantities called $B_{jk}$.
Clearly, the combinations occuring in \Eq{F} are very different from
those occuring in the mass matrix, \Eq{M}, because the two equations
involve different integrals and different combinations of the
couplings.  This implies that radiative decay amplitudes and
neutrinoless double beta decay can in principle provide independent
probes of CP violating phases.

\section{Conclusions}
Our primary intention has been the derivation of the radiative Majorana
neutrino decay amplitude in the $R$-parity violating supersymmetric
framework, with a careful handling of the phases from different
sources that enter the analysis.  
Our \Eq{Gamma} is the most general expression for the
radiative decay amplitude involving the $\lambda'$ couplings.  In the
derivation, we have not taken recourse to any specific assumption
regarding the masses and mixings of fermions and sfermions.  We have
relied only on the general field-theoretic features, highlightling
some subtle aspects regarding the Majorana character of the neutrinos
which have not been clearly dealt with in the literature.  We have
shown, even in this general framework, that the radiative decay
amplitudes depend on the quark masses and on the mass-squared
differences of squarks.  Our general formula can be used for all sorts
of different realizations of fermion and sfermion masses and mixings.
We have provided an illustrative example in Sec.~\ref{s:ans}, where
we have pointed out that radiative decay amplitudes and
neutrinoless double beta decay can in principle provide independent
probes of CP violating phases.  We
do not discuss the numerical implications of our result, because they
are of the same order of magnitude as those obtained in earlier
papers, e.g., in Refs.~\cite{bkp,faes}. 

As mentioned in the Introduction, we have not included the
contributions from the $\lambda$ couplings.  In fact, it is trivial to
include these contributions.  The diagrams will now involve charged
leptons and sleptons in the internal lines, instead of quarks and
squarks.  Thus the contributions can be directly adapted from the
$\lambda'$-induced contributions by making suitable substitutions.

\paragraph*{Acknowledgements~: }  GB and HP acknolwedge hospitality at
CERN Theory Division as short term visitors during the initial stages
of the work.  HP was supported by US DOE under the grant
DE-FG02-04ER41291.  TJW acknowledges support from DOE grant
DE-FG05-85ER40226, and the CERN Paid Associates program.



\begin{thebibliography}{99}


\bibitem{rpar} G. Farrar, P. Fayet, Phys. Lett. B {\bf 76} (1978) 575; 
S. Weinberg, Phys. Rev. D {\bf 26} (1982) 287;
N. Sakai, T. Yanagida, Nucl. Phys. B {\bf 197} (1982) 533;
C. Aulakh, R. Mohapatra, Phys. Lett. B {\bf 119} (1982) 136.

\bibitem{bilinear} A partial list: \\
A.Y. Smirnov, F. Vissani, Nucl. Phys. B 460
(1996) 37; E. Nardi, Phys. Rev. D {\bf 55} (1997) 5772; T. Banks,
Y. Grossman, 
E. Nardi, Y. Nir, Phys. Rev. D {\bf 52} (1995) 5319; R. Hempfling,
Nucl. Phys. B {\bf 478} (1996) 3; H. Nilles, N. Polonsky, Nucl. Phys.
B {\bf 484} (1997) 33; C. Liu, Mod. Phys. Lett. A {\bf 12} (1997) 329;
B.~Mukhopadhyaya, S.~Roy and F.~Vissani,
  Phys.\ Lett.\ B {\bf 443} (1998) 191
  [arXiv:hep-ph/9808265];
S.~Roy and B.~Mukhopadhyaya,
  Phys.\ Rev.\ D {\bf 55} (1997) 7020
  [arXiv:hep-ph/9612447].




\bibitem{trilinear} A partial list:\\
S. Dimopoulos, L. Hall, Phys. Lett. B {\bf 207} (1988) 210;
R. Godbole, P. Roy, X. Tata, Nucl. Phys. B {\bf 401} (1993) 67; 
M. Drees, S. Pakvasa, X. Tata
and T. ter Veldhuis, Phys. Rev. D {\bf 57} (1998) R5335; S. Rakshit,
G. Bhattacharyya, A. Raychaudhuri, Phys. Rev. D {\bf 59} (1999) 091701;
R. Adhikari, G. Omanovic, Phys. Rev. D {\bf 59} (1999)  073003; O. Kong,
Mod. Phys. Lett. A {\bf 14} (1999) 903; E.J. Chun, S.K. Kang, C.W. Kim,
U.W. Lee, Nucl. Phys. B {\bf 544} (1999)  89; 
K.~Choi, K.~Hwang and E.~J.~Chun,
  Phys.\ Rev.\ D {\bf 60} (1999) 031301
  [arXiv:hep-ph/9811363];
A.~S.~Joshipura and S.~K.~Vempati,
  Phys.\ Rev.\ D {\bf 60} (1999) 111303
  [arXiv:hep-ph/9903435];
Y.~Grossman and H.~E.~Haber,
  Phys.\ Rev.\ D {\bf 59} (1999) 093008
  [arXiv:hep-ph/9810536];
  Y.~Grossman and S.~Rakshit,
  Phys.\ Rev.\ D {\bf 69} (2004) 093002
  [arXiv:hep-ph/0311310];
S.~Rakshit,
  Mod.\ Phys.\ Lett.\ A {\bf 19} (2004) 2239
  [arXiv:hep-ph/0406168].


\bibitem{abl} 
A.~Abada, G.~Bhattacharyya and M.~Losada,
  Phys.\ Rev.\ D {\bf 66} (2002) 071701
  [arXiv:hep-ph/0208009]; 
S.~Davidson and M.~Losada,
  JHEP {\bf 0005} (2000) 021
  [arXiv:hep-ph/0005080]; 
S.~Davidson and M.~Losada,
  Phys.\ Rev.\ D {\bf 65} (2002) 075025
  [arXiv:hep-ph/0010325];
A.~Abada, S.~Davidson and M.~Losada,
  Phys.\ Rev.\ D {\bf 65} (2002) 075010
  [arXiv:hep-ph/0111332].


\bibitem{babu_moha} K.S. Babu, R.N. Mohapatra, Phys. Rev. Lett. {\bf
64} (1990) 1705; K. Enqvist, A. Masiero, A. Riotto, Nucl. Phys. B {\bf
373} (1992) 95; E. Roulet, D. Tommasini, Phys. Lett. B {\bf 256}
(1991) 218.

\bibitem{barbieri} R. Barbieri, M. Guzzo, A. Masiero, D. Tommasini,
Phys. Lett. B {\bf 252} (1990) 251.   

\bibitem{bkp}
  G.~Bhattacharyya, H.~V.~Klapdor-Kleingrothaus and H.~P\"as,
  Phys.\ Lett.\ B {\bf 463} (1999) 77
  [arXiv:hep-ph/9907432].
 
\bibitem{faes}
  M.~Gozdz, W.~A.~Kaminski, F.~Simkovic and A.~Faessler,
  arXiv:hep-ph/0606077.

\bibitem{general}  See, e.g., the scenarios discussed in 
  N.~F.~Bell, M.~Gorchtein, M.~J.~Ramsey-Musolf, P.~Vogel and P.~Wang,
  arXiv:hep-ph/0606248; 
  S.~Davidson, M.~Gorbahn and A.~Santamaria,
  Phys.\ Lett.\ B {\bf 626} (2005) 151
  [arXiv:hep-ph/0506085].

\bibitem{review} For reviews, see
G.~Bhattacharyya,
  Nucl.\ Phys.\ Proc.\ Suppl.\  {\bf 52A} (1997) 83
  [arXiv:hep-ph/9608415];
H.~K.~Dreiner,
  arXiv:hep-ph/9707435;
G.~Bhattacharyya, 
  arXiv:hep-ph/9709395; 
P.~Roy,
  arXiv:hep-ph/9712520;
S.~Raychaudhuri,
  arXiv:hep-ph/9905576;
M.~Chemtob,
  Prog.\ Part.\ Nucl.\ Phys.\  {\bf 54} (2005) 71
  [arXiv:hep-ph/0406029];
  R.~Barbier {\em et al.},
  Phys.\ Rept.\  {\bf 420} (2005) 1
  [arXiv:hep-ph/0406039].

\bibitem{gb95}
G.~Bhattacharyya, J.~R.~Ellis and K.~Sridhar,
  Mod.\ Phys.\ Lett.\ A {\bf 10} (1995) 1583
  [arXiv:hep-ph/9503264];
G.~Bhattacharyya and D.~Choudhury,
  Mod.\ Phys.\ Lett.\ A {\bf 10} (1995) 1699
  [arXiv:hep-ph/9503263].

\bibitem{Pal:1981rm}
  P.~B.~Pal and L.~Wolfenstein,
  Phys.\ Rev.\ D {\bf 25} (1982) 766.


\bibitem{Kayser:1983wm}
  B.~Kayser and A.~S.~Goldhaber,
  Phys.\ Rev.\ D {\bf 28} (1983) 2341.


\bibitem{Kayser:1984ge}
  B.~Kayser,
  Phys.\ Rev.\ D {\bf 30} (1984) 1023.

\bibitem{Schechter:1981hw}
  J.~Schechter and J.~W.~F.~Valle,
  Phys.\ Rev.\ D {\bf 24} (1981) 1883
  [Erratum-ibid.\ D {\bf 25} (1982) 283].

\bibitem{Nieves:1981zt}
  J.~F.~Nieves,
  Phys.\ Rev.\ D {\bf 26} (1982) 3152.

\bibitem{Mohapatra:1998rq}
  See, e.g., Sec. 7.2.4 of 
  R.~N.~Mohapatra and P.~B.~Pal,
  ``Massive neutrinos in physics and astrophysics.'' Third edition,
  [World Scientific, (2004)].

\bibitem{Doi:1980yb}
  M.~Doi, T.~Kotani, H.~Nishiura, K.~Okuda and E.~Takasugi,
  Phys.\ Lett.\ B {\bf 102} (1981) 323.

\end{thebibliography}
\end{document}